\documentclass{article}
\usepackage{arxiv}

\usepackage{graphicx}
\usepackage{orcidlink}
\usepackage{natbib}

\usepackage{authblk}

\usepackage[%
style=base,
singlelinecheck=false,%
format=plain,%
labelformat=simple,%
labelsep=period,%
justification=justified,%
font=none,small,%
labelfont=bf,%
textfont=none%
]{caption}
\usepackage{subcaption}
\usepackage{cleveref}
\crefname{figure}{figure}{figures}
\Crefname{figure}{Figure}{Figures}
\crefname{table}{table}{tables}
\Crefname{table}{Table}{Tables}
\usepackage{hyperref}

\usepackage{siunitx}

\makeatletter
\def\@email#1#2{%
 \endgroup
 \patchcmd{\titleblock@produce}
  {\frontmatter@RRAPformat}
  {\frontmatter@RRAPformat{\produce@RRAP{*#1\href{mailto:#2}{#2}}}\frontmatter@RRAPformat}
  {}{}
}%
\makeatother

\newcommand{\Table}[1]{\begin{table*}
  \setlength{\tabcolsep}{8pt}
  \caption{#1}
  \begin{tabular}{@{}l*{15}{l}}}

\def\endTable{\end{tabular}\end{table*}}

\newcommand{\br}{}
\newcommand{\mr}{\hline}

\title{Experimental Investigation of Time Series Classification using a Self-Pulsing Microring Resonator Network}

\date{May 26, 2026}

\author[1,2]{\orcidlink{0009-0001-3195-8620} Alessandro Foradori}
\author[2]{\orcidlink{0000-0002-6587-2614} Alessio Lugnan}
\author[2]{\orcidlink{0000-0001-7316-6034} Lorenzo Pavesi}
\author[1]{\orcidlink{0000-0001-6259-464X} Peter Bienstman}

\affil[1]{Photonics Research Group, Department of Information Technology, Ghent University-imec, Technologiepark Zwijnaarde 15, Ghent, 9052, Belgium}
\affil[2]{Nanoscience Laboratory, Department of Physics, University of Trento, Via Sommarive 14, Trento, 38123, Italy}

\renewcommand{\headeright}{}
\renewcommand{\undertitle}{}

\begin{document}
\maketitle

\begin{abstract}
Photonic neuromorphic computing offers compelling advantages in power efficiency and parallel processing, but often falls short in realizing scalable nonlinearity and long-term memory. These limitations can be overcome by silicon microring resonator (MRR) networks. These integrated photonic circuits enable compact, high-throughput neuromorphic computing by simultaneously exploiting spatial, temporal, and wavelength dimensions. This work provides an in-depth study of of MRR networks for photonics-based machine learning (ML). We investigate the system's effectiveness on two widely used image classification benchmarks, MNIST and Fashion-MNIST, by encoding images directly into time sequences. In particular, we enhance the computational performance of a linear readout classifier within the reservoir computing paradigm through the strategic use of multiple physical output ports, diverse laser wavelengths, and varied input power levels. Moreover, we explore a single-pixel classification setting, where inference does not require digital memory, thanks to the inherent memory and parallelism of our MRR network.
\end{abstract}

\section{Introduction}

Within the evolving field of artificial intelligence (AI), neuromorphic computing systems are gaining attention as hardware platforms that emulate the distributed, nonlinear functions of the biological brain. Among these, photonics-based neuromorphic architectures emerge due to their inherent advantages. In fact, they offer high throughput, low latency, energy-efficient linear operations and, in general, high parallelism by exploiting multiplexing techniques across multiple optical degrees of freedom, such as the optical wavelength or polarization \cite{li2025photonics, brunner2025roadmap}. These properties make neuromorphic  photonic solutions not only interesting for accelerating large-scale artificial neural networks in data centers \cite{farmakidis2024integrated}, but also attractive for compact, low-power edge computing applications. Specifically, when information is natively encoded in the optical domain, as in optical communications \cite{masaad2024experimental,argyris2022photonic} and in a broad variety of optical sensing implementations \cite{qin2022metasurface,venketeswaran2022recent,jayawickrema2022fibreoptic}. 

However, optical computing systems face challenges in providing energy-efficient scalable nonlinear nodes \cite{brunner2025roadmap}, required to implement artificial neurons and, in general, to host powerful machine learning (ML) models. For example, hybrid nonlinear nodes for neuromorphic computing based on Optical-to-Electrical-to-Optical (OEO) conversion have been introduced \cite{shastri2021photonics}, at the cost of increased complexity and energy consumption, which hinder scalability to large numbers of artificial neurons.  In addition, several edge computing applications require efficient and low-latency time-dependent signal processing, which would benefit from volatile or easily resettable optical memory effects \cite{liu2025physical}. Although this is relatively easy to obtain for memory durations in the nanosecond scale or shorter, e.g. via optical delay lines \cite{vandoorne2014experimental}, it has traditionally been difficult for longer timescales, such as in the millisecond to second range typical of most time-dependent sensing applications. All-optical non-volatile memory has been introduced in integrated photonics by employing phase change materials (PCMs), and their application for neuromorphic computing has been proposed in several works \cite{zhou2024fabrication,pernice2012photonic, rios2015integrated}. However, PCM-based all-optical memory requires switching the PCM with optical pulses, introducing relatively high power losses and difficulties in resetting the memory states in large optical networks \cite{feldmann2019alloptical}.

Recently, we demonstrated that photonic integrated networks of silicon microring resonators (MRRs) can provide all-optical nonlinearity and volatile long-term memory \cite{biasi2024exploring,lugnan2025reservoir}, with high-throughput production of nonlinear dynamical representations of the input signal. These representations can be exploited for computationally efficient ML \cite{lugnan2025emergent} based on the \textit{reservoir computing} (RC) approach \cite{jaeger2004harnessing,tanaka2019recent}, where a recurrent neural network (on a photonic circuit in our case) is randomly initialized and only a linear readout is trained to carry out a time-dependent ML task.

MRRs are versatile and widely used photonic devices, whose applications for neuromorphic computing have been proposed in several research works and diverse ways \cite{biasi2024photonic,tait2016microring,bazzanella2022microring, lugnan2025reservoir, mesaritakis2013micro, vaerenbergh2012cascadable, feldmann2019alloptical}. Traditionally, MRRs function as wavelength-selective filters, exploiting their resonance properties to control the spectral components of optical signals. In the context of neuromorphic computing, MRRs have often been proposed as linear nodes implementing optically weighted connections\cite{tait2016microring, shastri2021photonics, biasi2024photonic}. In this case, an MRR simultaneously extracts and weights a selected optical wavelength from a waveguide. 
Therefore, multiple MRRs can be applied to a single waveguide to implement the weighted multi-channel input of an optical artificial neuron with improved scalability.
The weights assigned to MRRs can be controlled by tuning their resonance conditions. This can be achieved through several mechanisms, such as via a local heater exploiting the thermo-optic effect, or using a integrated pn-junction for electro-optic modulation, or incorporating PCMs.

In addition, a silicon MRR can easily be driven in a nonlinear regime by raising the power of the resonant input light to the milliwatt range\cite{vaerenbergh2012cascadable,mancinelli2014chaotic}. Such a nonlinearity is due to the free carriers and heat accumulated in the silicon ring waveguide (mediated by the two-photon absorption effect), which change the ring refractive index, in turn producing a shift in the MRR resonant wavelength. Interestingly, the nonlinear effect based on free carriers blueshifts the MRR resonance, and is faster (lifetime around \SIrange{1}{45}{\nano\second}) but weaker than the thermo-optic nonlinearity (lifetime around \SIrange{60}{280}{\nano\second}), which produces a resonance redshift instead. It is important to note that, because of the different lifetimes of these two competing effects, a nonlinear MRR exhibits volatile memory with two different timescales. Although these effects are present in any silicon waveguide, in a MRR they are amplified and coupled together by its optical resonance, resulting in a much more energy-efficient and dynamic nonlinearity and memory\cite{lugnan2022rigorous, biasi2024photonic}. Thus, MRRs can be used as efficient photonic neurons. For example, the interplay between a MRR's free carriers and temperature in suitable conditions, gives rise to self-pulsing behavior, where a slowly-varying optical input is translated into a pulsed output signal\cite{biasi2024photonic}. Leveraging this property, several works have demonstrated that MRRs can be employed as artificial spiking neurons that are compact and simple to fabricate \cite{vaerenbergh2012cascadable, xiang2022alloptical, lugnan2022rigorous, biasi2024exploring, donati2025alloptical, xiang2025photonic}.

In this work, we expand the investigation of a MRR network for photonics-based ML, by testing our system on two popular image classification benchmark tasks, leveraging the multiple responses of the system across different wavelength and spatial ports. It should be stressed that, even if the MRR network is fixed after fabrication, it can produce a large number of different data representations, by varying input wavelength, power and output port. Therefore, the user can easily explore such a configuration space (given that RC systems allow for very fast and efficient training) and select those representations that are most suitable to learn a target ML task, which can be seen as a sort of \textit{train-to-train} procedure. In addition, we study the case where the classification is performed from a single time sample, which eliminates the need for digital memory. Finally, our system and the related performances are compared with those of comparable digital ANNs, to quantify the advantages presented by the proposed approach over the conventional ML systems. 
Specifically, the original contributions of this work present the following differences w.r.t. our previous works on MRR networks\cite{lugnan2025emergent, lugnan2025reservoir}:
\begin{itemize}
    \item Although it appears similar, the MRR network employed here is physically and functionally quite different w.r.t. the one employed in \cite{lugnan2025emergent}, as its long-term memory does not originates from lossy PCM but, in contrast, it comes from complex dynamics due to coupled self-pulsing in multiple MRRs (not present in \cite{lugnan2025emergent}).
    \item As opposed to \cite{lugnan2025emergent}, here we investigate how the ML accuracy changes by varying multiple key parameters, namely: the number of output network ports, the encoding modality and the input sample rate. Moreover, we consider an additional ML task (Fashion MNIST in addition to MNIST) and we study the network's functional memory by employing a single time sample per network output for ML.
    \item The full ML datasets are here employed and with minimal preprocessing, which is easily obtained without using digital processing but just by means of an optical modulator.  
    \item Comparison with digital ANN counterparts reaching similar accuracies, in terms of computational cost.
    \item In \cite{lugnan2025reservoir}, the same MRR network considered here is employed to demonstrate a fundamentally different ML approach, i.e. reservoir computing with non-volatile memory, aimed at exploiting the network's output much after the end of the input signal insertion. For this purpose, basic ad-hoc ML tasks were tackled, much simpler than the ones considered in this work, which is instead based on conventional reservoir computing.
\end{itemize}
Moreover, the scope of the mentioned previous works was to demonstrate novel systems and ML approaches. This article instead, given the growing interest in nonlinear MRR for neuromorphic photonics, aims to provide the community with a robust and in-depth analysis of the ML capabilities of these devices, when coupled together into a network.

The rest of this paper is structured as follows. 
\Cref{sec:principles}, \nameref{sec:principles}, describes the physics of the optical system, specifically its response to a constant signal (\cref{sec:self-pulsing}) and explores how this behavior can be exploited as a reservoir for machine learning tasks like classifying images encoded as time-dependent signals (\cref{sec:time-dependent-signals}).
This section also discusses the dimensionality expansion available in space, wavelength, and power (\cref{sec:dimensionality_expansion}), as well as the considered machine learning tasks (\cref{sec:ml-tasks}) and the specifics of input signal encoding, including sampling rate and power offset (\cref{sec:input_signal_and_data_acquisition}).
\Cref{sec:results}, \nameref{sec:results}, presents our findings, covering classification using full time-series representations across multiple ports (\cref{sec:multiport}), including a comparison with digital ANN, and the classification using single-pixel representations with optical memory (\cref{sec:single_pix}).
Finally, the \nameref{sec:conclusion} \cref{sec:conclusion} summarizes our work, while the \nameref{sec:methods} \cref{sec:methods} provides a detailed description of our experimental setup and procedures.

\section{Principles of Operation}\label{sec:principles}

Reservoir Computing (RC) is a machine learning paradigm where a fixed, untrained recurrent neural network transforms an input signal into a high-dimensional representation, which is then fed to a trained linear regressor or classifier.
It is particularly suitable for hardware implementation, since any complex enough, nonlinear and dynamical physical system can be employed as a reservoir, confining digital computation and training to a computationally cheap readout linear combination.

To function effectively as a reservoir, a physical system must exhibit nonlinearity, memory, and have the ability to expand the dimensionality of its input \cite{tanaka2019recent}.
In this context, the network of coupled microring resonators (MRRs) proposed here meets these requirements. 
Its nonlinear behavior, memory,  multiple accessible output ports and wavelength multiplexing capability make it a promising candidate for a compact integrated photonic reservoir with high throughput \cite{mesaritakis2013micro,lugnan2025reservoir}. 

\begin{figure}
    \centering
    \includegraphics[width=.5\linewidth]{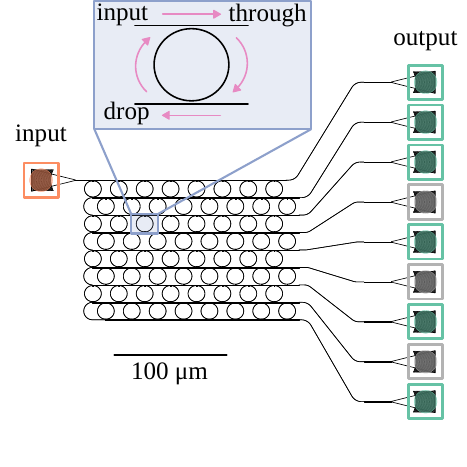}
    
    \caption{%
    \textbf{Schematics of the MRR network.}
    The network consists of 64 silicon microring resonators, coupled by straight waveguides and linked to one optical input port (left) and multiple output ports (right), consisting of grating couplers. 
    The output ports used for signal acquisition (ports 1, 2, 3, 5, 7, and 9, counted from top to bottom) are highlighted in green color; the remaining grey ports were not used due to low signal-to-noise ratio. The inset shows that light enters the microring resonator from the input port, passes either directly to the through port or couples into the ring and exits at the drop port, depending on whether the input optical frequency matches the resonant frequency.
    }
    \label{fig:show-schematics}
\end{figure}

Our system consists of an 8x8 matrix of coupled MRRs (\cref{fig:show-schematics}, details in \cref{sec:methods_pic}).
Each MRR is coupled to an upper and lower straight waveguide, which is terminated by grating couplers. These function as input or output ports, as they allow to insert and extract light via optical fibers.
In particular, the grating coupler on the left in the schematics of the circuit in \cref{fig:show-schematics} is used as input, while the ones on the right are used as output.

A single MRR acts as a filter around a resonant frequency, exhibiting a Lorentzian shape in its spectrum.
Ideally, when coherent light is injected into the waveguide on the top left side of a ring (input port), part of it can pass directly through to the top right side (through port) or couple into the ring and exit at the bottom left side (drop port), depending on how closely the optical frequency matches the MRR resonant frequency.
In particular, if the light is out of resonance, it will not couple with the MRR, and thus will continue unperturbed to the through port.
On the other hand, the more the incoming light is resonant, the more it will be redirected backwards through the opposite straight waveguide (part of it will be absorbed by the MRR) and the less it will proceed past the MRR (we refer the reader to \cite{biasi2024photonic} for more in-depth explanations).

The resonant frequency of an MRR is fundamentally determined by its material properties and geometric parameters, such as the effective refractive index and the optical path length.
However, the resonance exhibits sensitivity to temperature variations because of the thermo-optic effect, inducing changes in the refractive index. 
We control the temperature with a Peltier cell, which stabilizes the chip temperature in a range of \SI{0.1}{\kelvin}.
Additionally, fabrication tolerances introduce slight differences in the optical path and thus in the resonant frequency of MRRs, even when they are designed to be nominally identical.
Each ring resonance is thus fixed by construction but is randomly different from the others.
To mitigate the impact of these variations, we designed MRRs with relatively low quality factor, i.e. with a broader resonance linewidth, to ensure spectral overlap, and thus coupling, between different rings (\cref{fig:show-spectrum}). 

It should be stressed that the scalability of this type of MRR network in terms of number of coupled nodes (ensuring output readability) depends on the nonlinear dynamics originated by a certain optical input. Indeed, thanks to the nonlinear effects in MRRs, the ring resonances hindering a given optical path can be moved by a powerful enough input so that a readable output is obtained at an output port. For this reason, it is difficult to predict how large employable MRR networks can be. In addition to the effective network topology at a certain wavelength and input power, key factors are: the Q factors of the MRRs (resonnances of MRRs with higher Q factor are more easily "moved" by an optical input); the sensitivity of the employed photodetectors; and the time chracteristics of the input signal. Nevertheless, we expect that MRR networks quite larger (e.g., with double size) than the one here employed could be used with similar MRR Q factors and input power levels, as it is also evidenced in a previous work \cite{lugnan2025emergent}. Beyond this, higher MRR Q factors can enable the employment of much larger networks (or combinations thereof), as well as the usage of on-chip optical amplifiers.

\subsection{Response of the MRR Network to a Constant Signal}\label{sec:self-pulsing}

A sufficiently high input optical power (above roughly \SI{1}{\milli\watt}) temporarily and significantly affects the resonance frequency of an MRR via silicon nonlinear effects, due to changes in free carrier density and temperature of the ring waveguide \cite{biasi2024photonic}.
In an MRR network, high enough power can excite these nonlinear states in multiple rings simultaneously \cite{mancinelli2014chaotic}. 
Crucially, this phenomenon depends on the input power and on the detuning between the optical frequency of the input signal and the MRRs' resonance frequency.
Moreover, different nonlinear responses can also be observed at different output ports, as the light follows distinct propagation paths and couples to different resonators along the network.
Therefore, by employing different output ports and by varying the input laser power and optical frequency, our system can produce a large number of different nonlinear transformations of its input. Then, a suitable subset of representations can be selected to tackle a target ML task.

When a constant input is applied to the system, the output does not necessarily remain steady. 
Instead, the system can exhibit dynamic behaviors characterized by non-constant outputs.
Depending on the specific input condition and system parameters, the output can show periodic oscillations or more complex, non-periodic fluctuations \cite{biasi2024exploring}.
These dynamic responses are referred to as self-pulsing.

\begin{figure*}
    \centering
    \includegraphics[width=\linewidth]{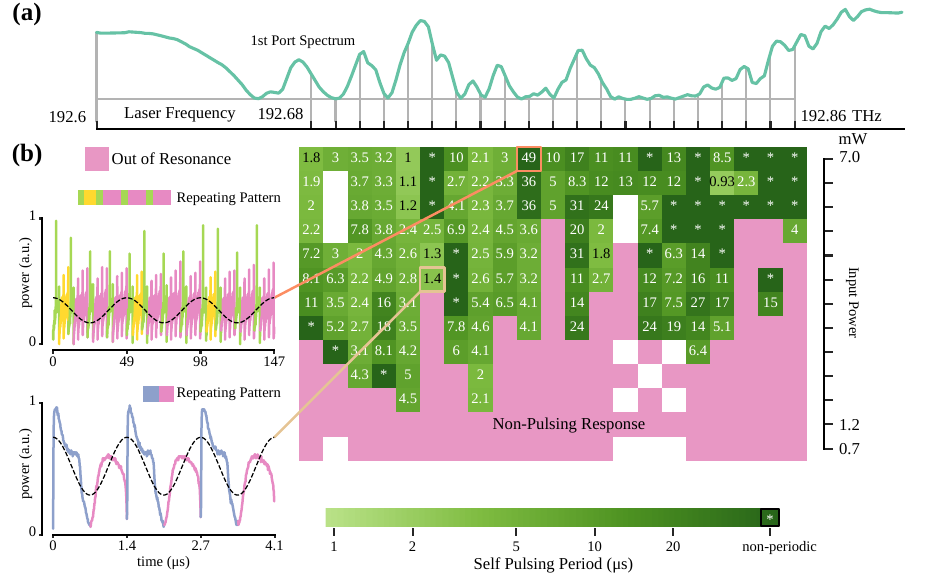}
    
    \begin{subcaptiongroup}
        \phantomcaption\label{fig:show-spectrum}
        \phantomcaption\label{fig:show-self-pulsing}
    \end{subcaptiongroup}
    
    \caption{%
    \textbf{Response to a Constant Signal.}
    \subref{fig:show-spectrum} Spectrum of port 1 acquired at low input power, in the absence of nonlinear effects.
    The x-axis (laser frequency) aligns with that used in the self-pulsing map.
    \subref{fig:show-self-pulsing} Color-map of the self-pulsing estimated period at the first output port of the network, as a function of input power and optical (laser) frequency. 
    The pink region indicates a response without self-pulsing, while green tones correspond to different self-pulsing periods.
    Cells marked with \(\ast\) denote those responses for which we could not detect a periodicity (see \cref{sec:methods-self-pulsing} for details).
    Two representative signals, with periods of \SI{49}{\micro\second} and \SI{1.4}{\micro\second}, are displayed on the left.
    Each signal is colored according to its repeating pattern, which visually represents the estimated period (also indicated by a dashed sinusoid). 
    }
    \label{fig:show-constant}
\end{figure*}

In \cref{fig:show-self-pulsing}, the estimated period of the self-pulsing oscillations at the first output port of the system is shown as a function of the constant input power and optical frequency, in those cases where some periodicity could be detected (see \cref{sec:methods-self-pulsing} in the Methods for further details).
The variety of detected periods spanning over a relatively broad range (from around 1 to tens of microseconds), highlights the richness and diversity of the network's response.

\subsection{Classification of Images Encoded to Time-dependent Signals}\label{sec:time-dependent-signals}

\begin{figure*}
    \centering
    \includegraphics[width=\linewidth]{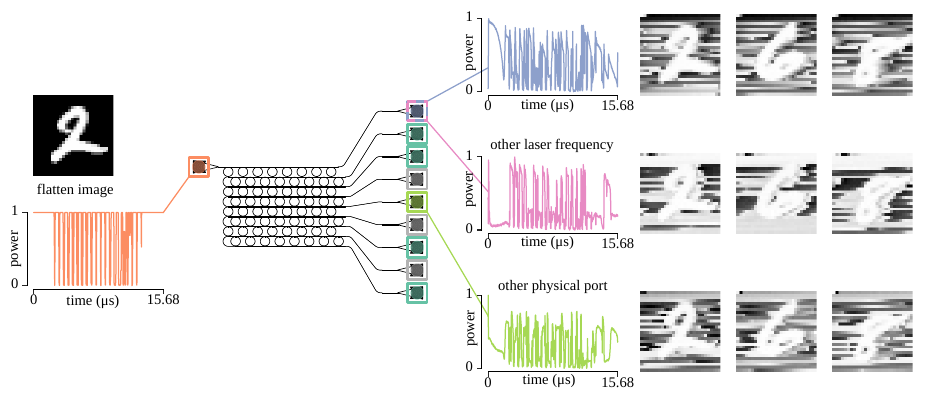}
    
    \caption{%
    \textbf{Multiple Representations of Images Encoded to Time-dependent Signals.}
    Each MNIST digit is first converted into a time sequence and then processed optically through the reservoir. The resulting temporal responses vary depending on the optical frequency and output port. These distinct output signals are reconstructed into images, illustrating the diversity of representations generated by the system.
    }
    \label{fig:show-example}
\end{figure*}

This variety of responses can be exploited for practical purposes, as we show here.
In particular, the system’s nonlinearity and memory make it suitable for RC.
For example, as shown in \cref{fig:show-example}, these dynamics can be used to increase the dimensionality of time-dependent input sequences. 
A handwritten digit from the MNIST dataset (a popular ML benchmark dataset comprising 10 classes of \(28 \times 28\) images \cite{lecun2010mnista}) is encoded as a time sequence and injected into the MRR network (more details can be found in \cref{sec:optical-setup} in the Methods). 
The resulting output time sequences provide diverse nonlinear representations (with memory) of the input sequence, thus performing a dimensionality expansion.
This transformation projects the input into a high-dimensional space, where features could become more easily linearly separable.
Such behavior aligns with the principle of reservoir computing, where the system dynamics create a fixed recurrent network and only a simple trainable readout layer is required for classification.

\subsection{Dimensionality Expansion}\label{sec:dimensionality_expansion}

Leveraging the system's capabilities to produce different responses, the same input signal can be processed under different conditions.
In particular, we process the input by acquiring outputs from different ports, at multiple optical frequencies and input power levels, in order to map it to a higher-dimensional space.

\paragraph{Physical Ports}
The output signals are measured at ports 1, 2, 3, 5, 7, and 9, counting from top to bottom on the right side w.r.t. \cref{fig:show-schematics}, while the other output ports present a too low signal-to-noise ratio.
In principle, measurements from all selected ports could be acquired simultaneously.
However, due to practical limitations in the current experimental setup, these measurements are performed sequentially.

\paragraph{Optical Frequencies}
Different optical frequencies, resonant with some MRRs in the network, are employed for signal acquisition. 
In particular, we use 10 equally spaced frequencies (unless otherwise stated) ranging from \SI{192.68}{\tera\hertz} to \SI{192.86}{\tera\hertz}. 
This frequency range corresponds to a region where the spectral content at port 1 is significantly rich, as shown in \cref{fig:show-spectrum}. 

\paragraph{Input Power Levels}
The system is also operated at different input power levels to modulate the network's nonlinearity. Because of the high complexity of the dynamics, the network's output signals can significant change even for moderate variations of input power, as can be appreciated in \cref{fig:show-self-pulsing}.
Average on-chip input powers are varied within the range of \SIrange{0.7}{7.0}{\milli\watt} for self pulsing characterization, and within \SIrange{1.1}{5.5}{\milli\watt} for the machine learning task. 
For machine learning applications, 5 (unless otherwise stated) equidistant power levels are employed, allowing the system to explore a wide set of dynamic behaviors and enhancing its processing versatility.

\paragraph{Baseline at Out-of-Resonance Frequency}
An additional measurement on the first output port, with optical frequency \SI{192.60}{\tera\hertz} and maximum input power, is performed as a baseline reference.
At this frequency, the light is not coupled to the MRRs and is directly conveyed to the first output port with negligible nonlinear effects. Therefore, this measurement allows to check what is the achieved ML performance when the MRR network is not used, thus providing a direct estimate of the advantage brought by our photonic system.
For this out-of-resonance measurement, the maximum input power is employed (in this case the response is linear and thus does not depend on the power level), in order to maximize the output SNR.

\subsection{ML Tasks}\label{sec:ml-tasks}
In this work, we focus on the classification tasks provided by the MNIST and Fashion MNIST datasets, which are a collection of small-size grayscale images (\(28 \times 28\) pixels) of handwritten digits and of clothes respectively, each comprising 10 classes \cite{lecun2010mnista,xiao2017fashionmnist}.
The training set includes 60,000 images, while the test set contains 10,000 images, following the standard train/test split conventionally adopted.
The choice of these datasets is motivated by several complementary considerations.

First, they constitute widely recognized benchmarks in the machine learning community and in the emerging field of neuromorphic photonics. Moreover, the transformation operated by the physical system on the input data is directly visualizable.

Indeed, since the raw temporal sequences generated by the reservoir are abstract and difficult to interpret, the spatial structure of the input data is leveraged to reconstruct the processed signals back into the image domain.
This approach allows for a qualitative appreciation of the nonlinear dynamics and memory effects introduced by the MRR network.

Second, the considered tasks are not a spatial pattern recognition problem. Each image is serialized into a one-dimensional time series by concatenating pixel values sequentially. Thus, the classification target is not an image but a time series, fully compatible with the recurrent, time-dependent processing paradigm of reservoir computing.

Third, while tasks such as parity (XOR of the previous N bits) and delayed memory recall (recovery of the Nth bit in the past) are standard benchmarks in the photonic reservoir computing literature, they probe a specific and relatively constrained form of memory. Image classification processed as a time series demands integration of information across 784 sequential inputs, exercising the memory of the reservoir in a richer and more complex way.

The MRR network operates as a photonic reservoir whose output is collected across different ports, frequencies and input powers.
These states are then fed to a linear classifier in software, which acts as a readout layer to perform the final classification.

\subsection{Input Signal and Data Acquisition}\label{sec:input_signal_and_data_acquisition}
To process the MNIST and Fashion-MNIST datasets in the photonic reservoir system, we first define an appropriate encoding scheme.
Both datasets consist of grayscale images with a resolution of \(28 \times 28\) pixels and 8-bit intensity levels, ranging from 0 (typically corresponding to the background in the image) to 255 (representing the strokes of the handwritten digits or of the dress).

\begin{figure*}
    \centering
    \includegraphics[width=\linewidth]{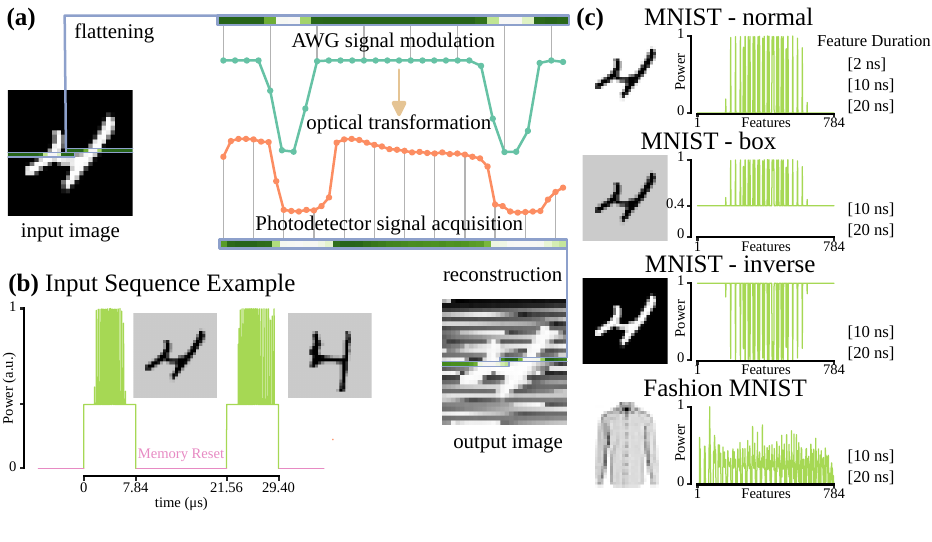}

    \begin{subcaptiongroup}
        \phantomcaption\label{fig:encoding-procedure}
        \phantomcaption\label{fig:encoding-input}
        \phantomcaption\label{fig:encoding-encodings}
    \end{subcaptiongroup}
    
    \caption{%
    \textbf{Encoding and processing of image data by the photonic network.}%
    \subref{fig:encoding-procedure} Overview of the data encoding and acquisition pipeline.
    A grayscale image from the MNIST or Fashion-MNIST dataset (28×28 pixels) is first flattened row by row into a 784-point sequence. This sequence is converted into a time series and used to drive an Arbitrary Waveform Generator (AWG), which modulates infrared laser light. The resulting optical signal is then injected into the MRR network, where it is optically processed. The output is detected by a photodetector and sampled by an oscilloscope (see \cref{sec:optical-setup} in Methods for details). To facilitate intuitive insight of the optical transformation, the acquired temporal signal is reshaped back into an image for a better visual comparison.
    \subref{fig:encoding-input} Time-encoded transmission of each image is followed by a brief pause (in the order of \SI{10}{\micro\second}, \cref{tab:osc_awg_timescale} in Methods), allowing the MRR network to return to its resting state. 
    This ensures that the system's response to each input is independent and unaffected by residual dynamics from previous samples.
    \subref{fig:encoding-encodings} List of combinations of encoding modes (normal, inverse, and box), features durations (i.e. the timestep of the AWG) and datasets (MNIST and Fashion-MNIST).
    }
    \label{fig:encoding}
\end{figure*}

Each image is flattened into a one-dimensional sequence of 784 points by reading pixel values row by row. This transformation produces a temporal sequence where each time sample corresponds to the intensity of a single pixel, with the encoding strategy determining the relationship between gray-scale intensity and optical output (\cref{fig:encoding-procedure}).
In particular, we employ the following encodings that are straightforward to achieve with a simple intensity modulator, without the need for further preprocessing:

\paragraph{Normal} A background pixel value of 0 corresponds to the minimum achievable optical power, while a value of 255 corresponds to the maximum achievable optical power. The background typically carries negligible optical power, and the nonlinear transformations only occur in the short pulses that correspond to the strokes of the image.

\paragraph{Inverse} This encoding inverts the image, such that a pixel value of 0 corresponds to the maximum optical power, and 255 corresponds to the minimum optical power. In this configuration, the strokes effectively act as interruptions of light insertion and, as a consequence, of light transmission through the MRR network.

\paragraph{Box} The minimum pixel value is mapped to \SI{40}{\percent} of the highest achievable optical power. Values between 0 and 255 are scaled accordingly within this restricted range. The background acts as a pump to the MRR network, possibily driving it into a nonlinear and self-pulsing state, which is then perturbed by the strokes.

\bigskip
Each feature (pixel) of the input sequence corresponds to a specific time step with fixed duration. 
In this work, we consider time step durations of \SI{2}{\nano\second}, \SI{10}{\nano\second}, and \SI{20}{\nano\second}, allowing us to explore which timescale is most effective for exploiting the self-pulsing behavior of the microring resonators.
These values are not chosen arbitrarily: the self-pulsing period of the device, operating in a non-chaotic regime, lies in the range of approximately \qtyrange{1}{50}{\micro\second}, and the total duration of a single input pattern (784 samples per image) spans \qtylist{2;10;20}{\nano\second} $\times$ 784 $\approx$ \qtylist{1.6;8;16}{\micro\second}, which is commensurate with this period.
Visual representations of the combinations of datasets, encodings and feature durations used are provided in \cref{fig:encoding-encodings}.

The complete input sequence is generated by modulating infrared laser light using an arbitrary waveform generator (AWG), which is programmed to continuously reproduce the entire dataset in a sequential loop.
Each image is transmitted as a time-encoded sequence, followed by a brief pause to allow the microring resonator system to relax and return to its nonexcited state before processing the next image (\cref{fig:encoding-input}). 
This pause ensures that the system’s response to each input remains independent and is not influenced by residual dynamics from previous samples.

The MRR network's output signal is received by a photodetector and converted into an electrical signal acquired by an oscilloscope. 
The sampling time of the oscilloscope is shorter than the one of the AWG, thus increasing the number of features per images of a factor of approximately 1.5 or 2.5, depending on the ratio between the AWG and oscilloscope sampling rate (see \cref{sec:data-acquisition} in the Methods for further details).

The reservoir dynamics associated with the best classification performance can differ across time steps, reflecting the fact that each time step selects a distinct operating point along the input modulation timescale. The time step acts as a hyperparameter of the reservoir system: while it can in principle be tuned, the optimal value is task-dependent and no universal optimum exists.

\section{Results}\label{sec:results}

In the first ML approach we consider in this work, we apply a linear readout classifier, as detailed in \cref{sec:multiport}, to different representations of the input time series (i.e. a flattened image) from different output ports, with fixed input power and frequency. 
Then, we present some comparisons with fully digital ANNs, to highlight the decrease in computational cost allowed by our photonic implementation.

In the second ML approach (\cref{sec:single_pix}), we select a single time sample (corresponding approximately to a pixel of the original image). We then combine all output representations of this single sample, gathered across various optical frequencies and output ports but fixing the input power level, and we feed them to a linear readout classifier.
This strategy leverages the intrinsic optical memory of the MRR network, thus removing the requirement for an external digital memory component.

In both cases, we evaluate the advantage brought by our photonic system via a comparison with suitable baseline results.

\subsection{Linear Readout on Full Time-Series Representations across Multiple Ports}\label{sec:multiport}

To evaluate the network’s computational capabilities, we encode each full flattened image as a time-domain signal and feed it into our MRR network. The system produces multiple parallel output signals, each of which represents a distinct feature projection of the input. 
These parallel representations can either be employed individually, by considering the signal from a single output port, or jointly, by combining signals across multiple ports. We assess classification performance under both scenarios, for fixed optical frequency and input power.
The aim of this analysis is to demonstrate how performance improvements are linked to both the internal dynamics of the photonic hardware and the combination of its representations. 
After measuring the responses of all ports, at first, each port's representation is independently used to train and test a linear classifier on the selected dataset.
Next, we combine the representations from multiple ports to assess the benefits of exploiting different optical paths as reservoir nodes.
Specifically in this case, we use 2 to 6 ports, and we explore different combinations of available representations (the first port is always included, to avoid that this exploration takes too much time).
It should be noted that not all the acquired output signals are included in this analysis, due to too low signal-to-noise ratio.
Moreover, it should be stressed that for this first ML approach, each classification is made considering output signals that can be easily acquired in parallel just by employing a single input laser frequency and using a photodetector for each output port. 

\subsubsection{Classification of hand-written digits: MNIST}

\begin{figure*}
    \centering
    \includegraphics[width=\linewidth]{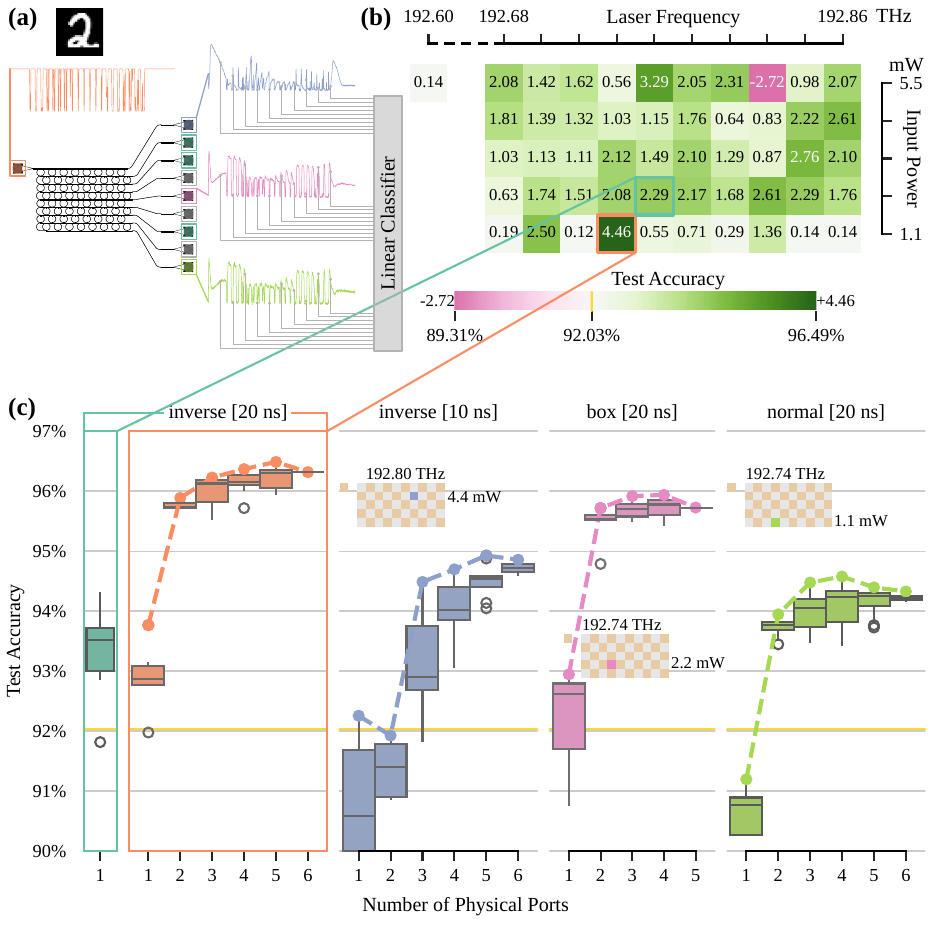}

    \begin{subcaptiongroup}
        \phantomcaption\label{fig:results-example}
        \phantomcaption\label{fig:result-heatmap}
        \phantomcaption\label{fig:result-boxplot}
    \end{subcaptiongroup}
    
    \caption{%
    \textbf{Classification performances on the MNIST dataset using representations from  multiple ports.}
    \subref{fig:results-example} Schematic example of the multiport classification strategy. Representations acquired from different output ports, under fixed optical frequency and input power, are stacked together and fed into a single linear classifier.
    \subref{fig:result-heatmap} Heatmap showing the best test classification accuracy obtained for each combination of laser frequency and input power, using the MNIST dataset with inverse encoding and a \SI{20}{\nano\second} AWG time step.
    For each pair, the number of ports used can range from one up to six (all available ports). If multiple ports are used, the first port is always included.
    The colormap indicates the test accuracy improvement over a baseline linear classifier trained on the raw (non-photonic) data, whose accuracy is \SI{92.03}{\percent}. 
    \subref{fig:result-boxplot} Boxplots showing test accuracy as a function of the number of combined ports, evaluated for different system configurations. 
    For each case, the optical frequency and input power providing the highest accuracy are chosen.
    A dashed line connects the highest accuracy (colored point) achieved at each port count, highlighting both the improvements given by combining multiple ports and the overfit that occurs with an excessive number of ports. Outliers are shown as empty points.}
    \label{fig:result}
\end{figure*}

In \cref{fig:result}, we present the best classification accuracies obtained for the MNIST dataset using a single linear classifier, while a complete set of final results for the different encodings is presented in \cref{tab:multiport_result_mnist}.
Each result is compared to a baseline, which is calculated by applying the same linear classifier directly to the original dataset, without any photonic processing (see \cref{sec:methods_results_software} in the Methods for more details).
The accuracy of this baseline model is \SI{92.03}{\percent}. 
For each combination of optical frequency and input power, we report in \cref{fig:result-heatmap} the highest accuracy achieved across all the output port combinations in the inverse configuration with \SI{20}{\nano\second} timestep.
In most cases, the photonic processing leads to a clear improvement in performance over the baseline.
The out-of-resonance configuration shows that a purely linear transformation by the employed photonic circuit (even though non-idealities of the employed modulator, optical amplifier and photodetector could introduce some simple nonlinearity and memory effects), i.e. without engaging the nonlinear and memory effects of the MRR network, does not provide a significant performance advantage (up to \SI{92.17}{\percent} accuracy).

The obtained accuracy improvements are therefore attributed to the MRR network’s dynamics, which perform a nonlinear mixing of the input features. This effect is particularly evident for the configuration (\SI{2.2}{\milli\watt}, \SI{192.76}{\tera\hertz}), where by using only one output port we achieve an accuracy of up to \SI{94.32}{\percent}. It should be stressed that, from the perspective of the data encoding, there is no dimensionality expansion provided by a single output port of the MRR network, since the number of temporal points (or pixels) is approximately the same for the input and for the output signals (for details, see \cref{sec:comparison}). However, the dynamical transformation performed by our photonic system spreads the information about the handwritten digit all over the background of the image, which would otherwise have a constant pixel value and thus carry no information in its intensity values (see examples in \cref{fig:show-example}). This effect corresponds to an effective dimensionality expansion of the information about the handwritten digit, thus resulting in a significant performance improvement.

In contrast, the small performance improvement (up to \SI{92.26}{\percent}) observed in the normal configuration, where the nonlinear mixing is confined to the short duration of the digit strokes, highlights the importance of the dynamics occurring in the background. This is evident when comparing it to the performance of the box (\SI{93.32}{\percent}) and inverse configurations, where the background effectively carries information.

Moreover, the additional dimensionality expansion provided by combining multiple ports contributes to further performance gains.
For instance, we achieve the best accuracy of \SI{96.49}{\percent} in the \SI{20}{\nano\second} inverse configuration with (\SI{1.1}{\milli\watt}, \SI{192.74}{\tera\hertz}), when combining signals from five ports.

\Table{{%
\textbf{Classification accuracy on the MNIST dataset using a single linear classifier applied to different combinations of output ports.}
The table presents the best test accuracy (4th column) achieved for a given configuration of encoding (1st column) and timestep duration (2nd column). 
The number of ports for which the maximum is obtained is shown in the 3rd column, while the best accuracy obtained using a single port, in the 5th column and the one in the out-of-resonance condition is in the 6th column.
The last row reports the test accuracy of the baseline linear classifier applied to the original (non-photonic) data. 
}
\label{tab:multiport_result_mnist}
}
    \br
     encoding & duration & \# ports & accuracy & single port & out-of-resonance\\
     \mr
     normal & \SI{2}{\nano\second} & 5 & \SI{94.50}{\percent} & \SI{92.29}{\percent} & \SI{91.54}{\percent}\\
     normal & \SI{10}{\nano\second} & 4 & \SI{94.82}{\percent} & \SI{91.77}{\percent} & \SI{91.57}{\percent}\\
     normal & \SI{20}{\nano\second} & 4 & \SI{94.58}{\percent} & \SI{92.26}{\percent} & \SI{91.78}{\percent}\\
     box & \SI{10}{\nano\second} & 3 & \SI{95.77}{\percent} & \SI{93.30}{\percent} & \SI{91.77}{\percent}\\
     box & \SI{20}{\nano\second} & 4 & \SI{95.94}{\percent} & \SI{93.32}{\percent} & \SI{91.80}{\percent}\\
     inverse & \SI{10}{\nano\second} & 5 & \SI{94.93}{\percent} & \SI{92.99}{\percent} & \SI{91.93}{\percent}\\
     inverse & \SI{20}{\nano\second} & 5 & \SI{96.49}{\percent} & \SI{94.31}{\percent} & \SI{92.12}{\percent}\\
     \mr
     \multicolumn{3}{@{}l}{Software LC} & \SI{92.03}{\percent} & & \\
     \br
\endTable

The boxplots in \cref{fig:result-boxplot} further highlight that these accuracy improvements are consistently observed across different combinations of ports, confirming the robustness of the multi-port strategy. 
Although similar improvement trends are observed when using the other encoding schemes or sampling times, the overall performance improvements in these cases are generally less pronounced.

Finally, we note that the decrease in the best accuracy observed in the plots of \cref{fig:result-boxplot} when using a higher number of ports is ascribed to overfitting.

\subsubsection{Classification of clothing items: Fashion-MNIST}
\Table{{%
\textbf{Classification accuracy on the Fashion-MNIST dataset using a single linear classifier applied to different combinations of output ports.}
The table presents the best test accuracy (4th column) achieved for a given configuration of encoding (1st column) and timestep duration (2nd column).. 
The number of ports for which the maximum is obtained is shown in the 3rd column, while the best accuracy obtained using a single port, in the 5th column and the one in the out-of-resonance condition, in the 6th column.
The last row reports the test accuracy of the baseline linear classifier trained on the raw (non-photonic) data. 
}
\label{tab:multiport_result_fmnist}}
    \br
     encoding & duration & \# ports& accuracy & single & out-of-resonance\\
     \mr
     normal & \SI{10}{\nano\second} & 4 & \SI{85.76}{\percent} & \SI{84.15}{\percent} & \SI{83.32}{\percent}\\
     normal & \SI{20}{\nano\second} & 4 & \SI{85.81}{\percent} & \SI{84.55}{\percent} & \SI{83.84}{\percent}\\
     \mr
     \multicolumn{3}{@{}l}{Software LC} & \SI{85.17} {\percent} & & \\
     \br
\endTable

The Fashion-MNIST dataset is another widely used benchmark in machine learning, offering a more challenging classification task compared to the original MNIST dataset. 
Like MNIST, it consists of 28×28 grayscale images, but instead of handwritten digits, it contains images of various clothing items, such as shirts, sneakers, and coats, each assigned to one of 10 classes.

Due to the increased complexity, Fashion-MNIST is significantly harder for simple models. 
For example, a linear classifier typically achieves a test accuracy of around \SI{85.17}{\percent} on Fashion-MNIST, compared to \SI{92.03}{\percent} on MNIST.

The approach relying on a single linear classifier and representations obtained from multiple ports yields only a modest improvement in accuracy (\SI{85.81}{\percent}, \cref{tab:multiport_result_fmnist}) over the baseline (\SI{85.17}{\percent}) of the fully software-based linear classifier.
This limited gain is ascribed to the noise introduced during the photonic transformation and the discretization constraints imposed by the oscilloscope's finite voltage resolution. 
These effects are further supported by the lower performance obtained using out-of-resonance representations, which reach at maximum \SI{83.84}{\percent} in accuracy.
Moreover, unlike MNIST, which consists mainly of high-contrast features (sharp edges between white background and dark strokes), Fashion-MNIST includes features distributed on a grayscale with smaller variation between each other.

\subsubsection{Comparison with Digital ANNs}\label{sec:comparison}

To contextualize the effectiveness of our neuromorphic photonic system, we compare the classification results presented in the previous section with those obtained using purely digital ANNs. 
This comparison serves not to establish 
competitive accuracy benchmarks, but rather to situate our photonic implementation within the broader landscape of neural network models of comparable architectural complexity. We stress that competing with digital neural networks on standard classification benchmarks 
is explicitly outside the scope of this work. The goal is instead to demonstrate that a 
physical nonlinear photonic substrate is capable of meaningful computation under realistic hardware constraints, with a number of digital linear and nonlinear operations lower than that required by conventional digital ANNs.

For the digital ANNs, we consider multilayer perceptrons (MLPs), which, like our photonic network, are not specifically structured to carry out the considered ML tasks.
Although convolutional neural networks (CNNs) generally achieve superior performance on image classification tasks when implemented in software, a direct comparison with CNNs would be misleading for two reasons. 
First, in this work the image data are presented to the photonic reservoir as temporally serialized sequences, thereby recasting what is nominally a spatial classification problem as a time series classification task. 
Second, a one-dimensional convolutional layer operating over a span of $N$ time steps is, in essence, a finite-memory function over the input features. In this sense, 1D-CNNs and reservoir computing systems share a common functional principle: both exploit a form of memory to encode temporal context into a fixed-dimensional representation. 
The distinction lies in whether that memory is implemented through learned convolutional filters or through the dynamics of a physical nonlinear system. 
Given this equivalence, the MLP represents a more appropriate and architecturally neutral reference for benchmarking purposes, as it imposes no implicit temporal structure on the input and thus reflects more faithfully the computational overhead required in the absence of task-specific inductive biases.

We evaluate the classification performance on the MNIST dataset.
In \cref{tab:software_mnist} we show the results obtained using different software machine learning architectures. The first one is a linear classifier (logistic regression, which is often used as a baseline test accuracy) trained with Stochastic Gradient Descent (SGD)\cite{alpaydin2020introduction}. The second one is 
a feedforward neural network composed of an input layer (784 features), a hidden layer with a variable number of neurons with a ReLu function and an output layer with 10 nodes\cite{alpaydin2020introduction}. The weights are trained using SGD (see \cref{sec:methods_results_software} in Methods).
For each configuration, we report the test accuracy along with the number of trainable parameters and the total number of software-based nonlinear operations required.
These metrics serve as a reference point for comparison with our photonic reservoir computing approach based on the MRR network.

We first establish a baseline using a simple linear classifier (logistic regression). This model, when applied to the original MNIST data, achieves an accuracy of \SI{92.03}{\percent}.
By processing the data with our MRR network using only a single port and then applying the linear classifier, we achieve a higher accuracy of \SI{94.32}{\percent}. Interestingly, this improvement was accomplished with only a slight increase in trainable parameters, due to the fact that we oversample the waveforms in the photonics case.

Then, we compare our system to conventional multilayer neural networks (ANN) with one hidden layer. First, we configure a digital ANN to have a similar number of trainable parameters as our multiport photonic system (approximately \num{60e3}).  This ANN, with 77 neurons in its hidden layer, reaches an accuracy of \SI{96.07(10)}{\percent}. In contrast, our system achieved a higher accuracy of \SI{96.49}{\percent} with fewer nonlinear operations.

Next, we trained a more complex digital ANN to match our system's accuracy of approximately \SI{96.5}{\percent}). To achieve this, the ANN required a hidden layer with 120 neurons, increasing the number of trainable parameters to \num{95e3} and its nonlinear operations to 130 (compared to just 10 for our system's softmax readout layer). This highlights an advantage of our photonic approach: it delivers the same accuracy with a lower number of trainable parameters and nonlinear software operations.

\Table{{%
\textbf{Best classification accuracy on the MNIST dataset, comparing photonic and digital network architectures.}
The first three rows show results from fully digital networks trained on the original data, while the last two rows present results from a photonic readout trained using a representation from the inverse representation with a 20 nanosecond timestep.
The first column specifies the network type.
The second column provides a detailed description of the network's architecture by listing the number of nodes for each layer (input - [hidden -] output). Note that the MRR network output is oversampled (1225 time samples) with respect to the original number of pixels (784). 
The remaining columns quantify the number of trainable parameters, the number of nonlinear operations, and the best test accuracy (eventually expressed in concise uncertainty notation, with the error given as the standard deviation over 10 repeated measurements; see \cref{sec:methods_results_software} in Methods) achieved for each model.
}
\label{tab:software_mnist}}
    \br
      & & {trainable} & {nonlinear} &  \\
system      & layers & {parameters} & {operations} & accuracy \\
     \mr
     Linear Classifier (LC) & 784 - 10 & 7850 & 10 & \SI{92.03}{\percent} \\
     Multilayer ANN & 784 - 77 - 10 & 61225 & 87 & \SI{96.07(10)}{\percent} \\
     Multilayer ANN & 784 - 120 - 10 & 95410 & 130 & \SI{96.53(7)}{\percent} \\
     \mr
     MRR net (1 output) + LC & 1225 - 10 & 12260 & 10 & \SI{94.32}{\percent} \\
     MRR net + LC & \((1225 \times 5)\) - 10 & 61260 & 10 & \SI{96.49}{\percent} \\
     \br
\endTable

\subsection{Classification using Single Pixel Representations with Optical Memory}\label{sec:single_pix}

\begin{figure*}
    \centering
    \includegraphics[width=\linewidth]{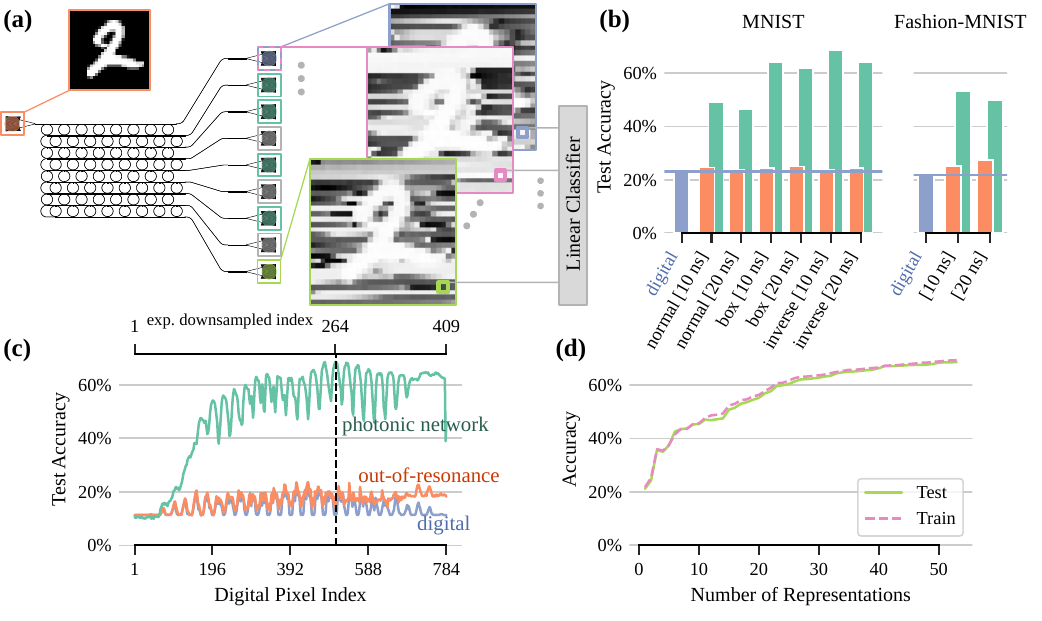}

    \begin{subcaptiongroup}
        \phantomcaption\label{fig:single_pix_example}
        \phantomcaption\label{fig:single_pix_summary}
        \phantomcaption\label{fig:single_pix_index}
        \phantomcaption\label{fig:single_pix_representations}
    \end{subcaptiongroup}
    
    \caption{%
    \textbf{Classification using a single time sample (or pixel).}
    \subref{fig:single_pix_example} A linear classifier is applied to a specific time sample (or image pixel) of the nonlinear representations generated by the MRR network. \subref{fig:single_pix_summary} Bar plot presenting the corresponding maximum test accuracies (photonic network, green bars) obtained for the considered classification tasks (x axis, see \cref{sec:input_signal_and_data_acquisition}). They are compared with their baseline counterparts (out-of-resonance, orange bars), obtained without processing from our MRR network (and completely digitally in the case of the blue bars). \subref{fig:single_pix_index} Pixel-by-pixel accuracy in the case with highest maximum classification accuracy, i.e. the MNIST classification with \textit{inverse} encoding and \SI{10}{\nano\second} time sample duration. \subref{fig:single_pix_representations} Corresponding classification accuracy, for the best performing pixel (with index 264), as a function of the number of employed representations.
    }
    \label{fig:single_pix}
\end{figure*}

In \cref{sec:multiport} we have shown that the nonlinear dynamics and memory provided by our MRR network can generate multiple representations of an input image, which are useful to improve the classification accuracy of a linear classifier. In this case, however, digital memory is still required for ML inference, since the linear classifier is applied to the recorded signals produced by the photonic network. In this section, we will instead study a case where no digital memory (of the signals to be processed) is required for ML inference, i.e. when the linear classifier is directly applied to the MRR network's outputs across various optical frequencies and spatial ports at a specific time, corresponding to a single pixel of the processed image (\cref{fig:single_pix_example}). It should be stressed that this demonstration is not about solving image classificaiton with high accuracy, but is meant to provide the reader with an idea of how much useful information can be conveyed by the MRR network output at a single time. This setting can be particularly useful when tackling simpler ML tasks requiring memory on an edge computing system with stringent restriction on computational complexity, energy consumption or latency. 

\Table{{%
\textbf{Classification accuracy using a single time sample (or pixel).}
The table presents the best test accuracy (4th column) achieved for a given dataset (1st column) and a given configuration of encoding (2nd column) and timestep duration (3rd column). 
The accuracy obtained in the out-of-resonance condition is shown in the 5th column.
}
\label{tab:single_pixel}
}
    \br
     task & encoding & duration & accuracy & out-of-resonance\\
     \mr
     MNIST & normal & \SI{10}{\nano\second} & \SI{49.30}{\percent} & \SI{24.58}{\percent}\\
     MNIST & normal & \SI{20}{\nano\second} & \SI{46.43}{\percent} & \SI{23.64}{\percent}\\
     MNIST & box & \SI{10}{\nano\second} & \SI{64.13}{\percent} & \SI{24.35}{\percent}\\
     MNIST & box & \SI{20}{\nano\second} & \SI{62.05}{\percent} & \SI{24.90}{\percent}\\
     MNIST & inverse & \SI{10}{\nano\second} & \SI{68.57}{\percent} & \SI{23.67}{\percent}\\
     MNIST & inverse & \SI{20}{\nano\second} & \SI{64.09}{\percent} & \SI{24.29}{\percent}\\
     \mr
     Fashion-MNIST & normal & \SI{10}{\nano\second} & \SI{53.13}{\percent} & \SI{25.16}{\percent}\\
     Fashion-MNIST & normal & \SI{20}{\nano\second} & \SI{50.01}{\percent} & \SI{27.36}{\percent}\\
     \br
\endTable

In particular, we consider all the available output representations for a given input optical power, we downsample them with a factor 3 to reduce experimental noise and computation time (see \cref{sec:methods_results_pixel} in Methods for details). At each moment in time, we use the different outputs of the network as features for the linear classifier. E.g., if we have 60 representations (10 laser frequencies per output port) at a given input power, our classifier will be applied to 60 values, each taken at the same time from the 60 parallel representations. We repeat this for all the available output pixels (i.e. the value corresponding to a single time sample of the downsampled output time series) and we select the one with highest validation accuracy (the dataset is split into 55000 training samples, 5000 validation samples and 10000 test samples). The test accuracies are subsequently computed using this selected index and are shown in \cref{fig:single_pix_summary} and reported in \cref{tab:single_pixel} for the different classification cases considered (i.e., different encodings of the MNIST and the Fashion-MNIST datasets, explained in \cref{sec:input_signal_and_data_acquisition}). An experimental baseline accuracy is computed for each case, considering a single out-of-resonance linear representation (discussed in \cref{sec:dimensionality_expansion}), corresponding to the case where no processing by the MRR network occurs. Moreover, a fully digital baseline accuracy is computed for the MNIST and Fashion-MNIST tasks using the original datasets. It should be stressed that, due to setup limitations, we performed sequential measurements of network responses at different input wavelengths. We consider the corresponding obtained ML accuracies as approximations of the accuracies we would obtain in the fully parallel case, where different wavelengths are injected concurrently and might influence each others in the MRR network.

While the baseline accuracies are usually below \SI{25}{\percent} (except for the Fashion-MNIST case, where it reaches \SI{27.4}{\percent}), by employing our photonic network we reach \SI{68.6}{\percent} for the MNIST (with \textit{inverse} encoding and pixel duration of \SI{10}{\nano\second}) and \SI{53.1}{\percent} for the Fashion-MNIST (pixel duration of \SI{10}{\nano\second}). Such an improvement is ascribed to the diverse memory effects provided by the different representations generated by the MRR network, where the value of a single pixel retains information about the previous image pixels. This conclusion is supported by the observed trend in classification accuracy as a function of the considered pixel index (in chronological order), shown in \cref{fig:single_pix_index}. The single-pixel accuracy notably increases with the pixel index (in addition to fluctuations due to the image spatial structure), as the network's memory builds up until a single pixel conveys enough information about a substantial part of the full image. As expected, this is clearly not the case for the experimental baseline accuracy, because of the absence of optical memory. 

Moreover, we investigate how the number of parallel nonlinear representations benefits single-pixel accuracy, for the best performing case (i.e., MNIST classification with \textit{inverse} encoding and \SI{10}{\nano\second} pixel duration). Specifically, we calculated the classification accuracy while increasing the number of employed representations (corresponding to the number of single-pixel features fed into the classifier), starting from the lowest laser frequency at the first output port and gradually adding more output ports and laser frequencies (\cref{fig:single_pix_representations}). The corresponding plot presents a continuously increasing trend, thus showing that most of the different nonlinear representations generated by our MRR network are useful to significantly improve the linear classifier's performance.

\section{Conclusion}\label{sec:conclusion}
We studied the highly complex and diverse nonlinear dynamics hosted by a microring resonator (MRR) network (namely a silicon photonic integrated circuit comprising 64 coupled MRRs) and its multi-wavelength parallelism for the classification of images encoded as a time-dependent optical signal. In particular, we investigated the capability of the MRR network to expand its input dimensionality through the use of multiple physical output ports, laser wavelengths and input power levels, in order to enhance the computational performance of a linear readout classifier (as per the \textit{reservoir computing} scheme). To do so, we considered two popular machine learning benchmark tasks on image classification, respectively from the MNIST (handwritten digits) and Fashion-MNIST (clothing items) datasets. By stacking together the signals from different output ports, we respectively achieved classification accuracies of \SI{96.49}{\percent} and \SI{85.81}{\percent}, starting from corresponding baselines (accuracy without photonic network) of \SI{92.12}{\percent} and \SI{83.83}{\percent}. 

This photonic approach provides an efficient alternative to traditional digital neural networks.
For instance, when using approximately \SI{60e3} trainable parameters, our system achieved a \SI{96.49}{\percent} accuracy on MNIST dataset. However, a standard digital ANN with the same number a trainable parameters only reached \SI{96.07(10)}{\percent}.
To match our system performance, a digital ANN would require a hidden layer with 120 neurons instead of 77, which would increase the number of trainable parameters to \num{95e3} and the number of nonlinearities to 130. In contrast, our system's readout linear classifier only requires 10 softmax operations to achieve the same accuracy.

Moreover, we studied the case where the readout classifier is applied to a single time sample (or pixel) rather than to the whole image, which in principle enables machine learning inference without digital memory. We obtained that the optical memory and dimensionality expansion performed by the MRR network allowed classification accuracies of \SI{68.57}{\percent} and \SI{53.13}{\percent} for the MNIST and Fashion-MNIST tasks respectively, starting from baselines of \SI{23.67}{\percent} and \SI{25.16}{\percent}. Although not directly relevant to the goal of accurate image classification, this result shows that the MRR network can greatly improve the performance of fast and computationally cheap digital models when tackling simple tasks that require working memory. This is significant, e.g., for edge computing applications with stringent restriction on computational cost, energy consumption or latency. 

Further advantages of our MRR network are compactness (\SI{0.15}{\milli\meter\squared}), low energy consumption (a few milliwatts of input optical power), and a technologically mature and CMOS-compatible platform. Furthermore, our system can be easily scaled up by employing more physical ports, MRRs and wavelength channels. Exciting future directions are direct application to photonic sensors producing time-dependent signals and making the network internally trainable by applying optical modulators to the MRRs.

\section{Methods}\label{sec:methods}

\subsection{Photonic Integrated Circuit}\label{sec:methods_pic}

We employ a photonic integrated circuit (PIC) fabricated by imec (Leuven, Belgium) on a silicon-on-insulator (SOI) platform.
The silicon waveguides have a cross-sectional dimension of \(\SI{450}{\nano\meter} \times \SI{220}{\nano\meter}\) and are embedded in a silica cladding.

The circuit comprises a network of 64 microring resonators coupled with bus waveguides, as in \cref{fig:setup-photo}.
Adjacent MRRs in the same row are spaced by \SI{22.7}{\micro\meter}. MRRs  are horizontally offset by \SI{11.35}{\micro\meter} compared to the neighboring rows.
 Each MRR features a racetrack geometry, with a bend radius of \SI{7}{\micro\meter} and straight coupling sections of \SI{0.71}{\micro\meter}.
The coupling gap between the bus waveguides and the resonators is \SI{0.2}{\micro\meter}.

The circuit includes one input port and nine output ports at the end of the bus waveguides, each equipped with a grating coupler with a nominal insertion loss of approximately \SI{3}{\decibel}.

\subsection{Experimental Setup}\label{sec:optical-setup}

\begin{figure*}
    \centering
    \includegraphics[width=\linewidth]{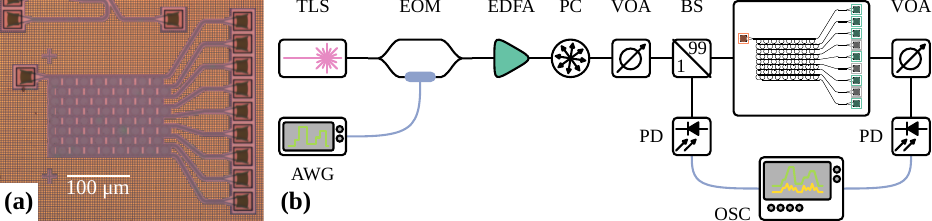}

    \begin{subcaptiongroup}
        \phantomcaption\label{fig:setup-photo}
        \phantomcaption\label{fig:setup-schematics}
    \end{subcaptiongroup}
    
    \caption{%
    \textbf{Micrograph of the MRR network and employed experimental setup.}
    \subref{fig:setup-schematics} The coherent light (wavelength around \SI{1550}{\micro\meter}) of a tunable laser source (TLS) is modulated by an electro-optical modulator (EOM) controlled by an arbitrary waveform generator (AWG). The resulting optical signal is then amplified by an erbium-doped fiber aplifier (EDFA) and passes through a polarization controller (PC) and a variable optical attenuator (VOA). Then, the optical beam is split by a \SI{99}{\percent}-\SI{1}{\percent} beam splitter (BS), where the smaller fraction of light power is read by a slow photodetector (PD) as an input power reference. The rest of the optical power is coupled via a cleaved fiber into the input directional coupler of the integrated MRR network. Similarly, the output signal is picked up by a second cleaved fiber and fed into a variable optical attenuator (VOA) before being acquired by a fast PD. Both PDs' electric outputs  are sampled and stored by an oscilloscope (OSC).
    }
    \label{fig:setup}
\end{figure*}

The optical setup (\cref{fig:setup-schematics}) used for working with integrated photonics consists of a transmission stage, in which the information is converted from an electric signal into an optical one, and a receiver stage, in which is converted back to an electric signal. The light travels inside the Photonic Integrated Circuit (PIC), or inside polarization-maintaining optical fibers, which link all the optic-related instruments required for the correct functioning of the system.

A continuous-wave tunable laser (TLS, Pure Photonics) provides optical input at a fixed power and selected wavelength, typically centered around \SI{1550}{\nano\meter}. 
This wavelength corresponds to an optical frequency in the C-band and an energy slightly above half the bandgap of silicon. 
As such, it is sufficient to induce nonlinear effects, such as two-photon absorption, that are essential to the dynamical behavior of the microring resonator network.

The optical signal is produced by modulating the laser light with an electro-optic modulator (EOM, iXblue model MXAN-LN-10), driven by an arbitrary waveform generator (AWG, Spectrum model DN2.663-02).

Then, it is amplified with an erbium-doped fiber amplifier (EDFA, Thorlabs) and adjusted to the desired power level using an electronic variable optical attenuator (VOA).

Light, in addition to its intensity, has other degrees of freedom. 
One of the most important ones is polarization, which describes the orientation of the electric field vector of the wave. 
Photonic components are typically sensitive to polarization and are optimized for a specific polarization state. To ensure efficient coupling and consistent behavior of the microring resonator network, a polarization control stage (PC) is used to adjust the input light's polarization.

We monitor the average input power by directing about \SI{1}{\percent} of the signal to a slow photodetector (PD, 30 kHz bandwidth, New Focus model M2033), splitting the light with a 99:1 fiber splitter.
The remaining part is inject into the PIC via a fiber.
The terminal side of the optical fiber is mounted on a three-axis linear piezoelectric stage and inclined with a specific angle (about \SI{10}{\degree} w.r.t. the normal of the PIC).
The light which exits the fiber is coupled to the input grating coupler.
Analogously, the transmitted light is sent back to another optical fiber from an output grating coupler.
Finally, the signal is collected by a fast photodetector (PD, 600 MHz, Menlosystem model FPD610-FC-NIR), which is preceded for safety by a second VOA.
This second VOA safeguards the fast photodetector from excessively high optical power.

The electrical signals from both photodetectors are acquired by an oscilloscope (OSC, Picoscope model 6000).
The chip temperature is actively stabilized by a thermostat system controlled by a PID controller connected to a Peltier cell and a \SI{10}{\kilo\ohm} thermistor.

\subsubsection{Data Acquisition}\label{sec:data-acquisition}

Each dataset is encoded as an optical time sequence, as previously described in \cref{sec:input_signal_and_data_acquisition}. Specifically, the sequence consists of flattened images interleaved with short breaks that reset the system to its initial state. A longer pause is inserted at the end to mark the sequence boundary. This entire representation is repeatedly generated by the AWG.
The available AWG time steps are constrained by the discrete set of sampling rates supported: the AWG operates at a maximum rate of \SI{500}{MSa\per\second}, with only a finite number of lower rates available.
The AWG and oscilloscope operate at different, not synchronized sampling rates, as detailed in \cref{tab:osc_awg_timescale}.
To ensure adequate temporal resolution, we opted for a slightly denser sampling for the output signal compared to the input one, so as to avoid excessive oversampling, which would rapidly fill the acquisition memory and limit the recordable sequence length.

\Table{{%
\textbf{Sampling time of the employed instrumentation used to generate and acquire the time sequences.}
The sampling time of the Arbitrary Waveform Generator (AWG) and the oscilloscope (OSC) are shown in the first two columns. The ratio between these two timesteps is provided in the 3rd column, which also corresponds to the upsampling ratio from the original 784 features to the final number of features (4th column). The duration of a single image and the brief pause between two consecutive images are displayed in the 5th and 6th columns, respectively.
}
\label{tab:osc_awg_timescale}}
    \br
     AWG & OSC & ratio & \# features & duration & pause \\
     \mr
     \SI{2}{\nano\second} & \SI{0.8}{\nano\second} & 2.5 & 1960 & \SI{1568}{\nano\second} & \SI{2744}{\nano\second} \\
     \SI{10}{\nano\second} & \SI{6.4}{\nano\second} & 1.5625 & 1225 & \SI{7.84}{\micro\second} & \SI{13.72}{\micro\second} \\
     \SI{20}{\nano\second} & \SI{12.8}{\micro\second} & 1.5625 & 1225 & \SI{15.68}{\micro\second} & \SI{15.68}{\micro\second} \\
     \br
\endTable

The following procedure is carried out for each available output port.

The data acquisition begins with an initial sequence captured at an out-of-resonance laser wavelength and maximum input power, aiming to maximize the signal-to-noise ratio. At this stage, the system operates within a linear regime, ensuring that the measured signal is only subject to linear transformations.

Subsequently, the system's response is recorded across a range of resonant laser frequencies and input power levels. The outer acquisition loop varies the laser frequency, while for each selected frequency, the input power is adjusted using the VOA.

During this process, the oscilloscope records a signal that lasts slightly more than twice the length of the full sequence. This ensures that at least one full, uninterrupted representation of the dataset is captured.

Finally, a second out-of-resonance sequence is recorded under the same conditions as the initial one. This final measurement verifies the stability of the system and ensures that no significant changes occurred during the acquisition, thus validating the reliability of the collected data.

\subsection{Estimation of Self-Pulsing Regime}\label{sec:methods-self-pulsing}

The discrimination between self-pulsing and constant output is done by computing the sum of the absolute value of the Fourier transform of the acquired signals.
This scalar feature separates the two regimes, forming two distinct clusters that do not overlap, making them linearly separable.
The self-pulsing period is then computed as the lag greater than zero corresponding to the highest peak in the autocorrelation.
If there is no peak or if the peaks are lower than 0.5, then the series is considered non-periodic.
The autocorrelation is computed via the Wiener–Khinchin theorem, since the Fourier transform is already available. We believe it likely that the self-pulsing states without a period estimate (the ones with an asterisk in the colormap in \cref{fig:show-self-pulsing}) are indeed chaotic states, although a more in-depth analysis (beyond the scope of this work) is required to confirm this. Other possible but unlikely reasons for a missing period estimate are that our measured waveform was not long enough to detect periodicity or that the signal-to-noise ratio was too high.

\subsection{Machine Learning Aspects}

\subsubsection{Linear Classifier (LC)}\label{sec:methods_results_multiport}
For baseline comparison, we utilize a linear classifier (LC) implemented in \textit{Tensorflow} and consisting in a single \textit{Dense} layer.

The classifier's input size corresponds to the number of features. The ten classes were represented using a one-hot encoding scheme.

The model is trained using the \textit{CategoricalCrossentropy} loss function, with the parameter \textit{from\_logits \(=\) True}. This setting enables the loss function to perform the final softmax operation internally, which enhances computational efficiency and numerical stability.

Model parameters are updated using Stochastic Gradient Descent (\textit{SGD}).
A batch size of 64 is used for training.

Input data is normalized using the training set to achieve a zero mean and a unit standard deviation

Due to the convex nature of its loss landscape, a properly regularized linear classifier consistently converges to the same accuracy, which is why a single training run is sufficient for its evaluation.

\paragraph{Digital}
The LC classifier is applied to the normalized original data.

\paragraph{Single Port}
The output signal from the MRR network serves as the input for the LC. A \textit{Dropout} layer is placed after the input representation and before the \textit{Dense} layer, which reduces the feature space to 784 during training. 
The chosen \textit{Dropout} rate acts as a regularization mechanism, and its value was chosen through hyperparameter optimization, which tunes it over a validation set.
This approach also mimics the downsampling performed on an oscilloscope, where the signal is oversampled to mitigate noise and subsequently averaged to extract a single, representative value per time step.

\paragraph{Multiple Ports}
Similar to the single-port approach, each representation is reduced to 784 features during training using a \textit{Dropout} layer. These features are then concatenated to form a single input for the LC. Therefore, the total number of input features for the LC is the product of the number of features per port and the number of ports.

\subsubsection{Multilayer ANN in Software}\label{sec:methods_results_software}
In contrast to the linear classifier, the training of a multilayer neural network can lead to different final solutions depending on the initial conditions. To mitigate this variability and obtain a more robust performance estimate, we trained the multilayer ANN 10 times with different random weight initializations. The results presented are the average of the accuracies obtained from these 10 training runs, along with their standard deviation.

Our multilayer ANN is relatively simple, consisting of an input layer followed by a single \textit{Dense} hidden layer with a variable number of neurons and a ReLU activation function. A \textit{Dropout} layer with a rate of 0.5 is applied after this hidden layer for regularization. The network concludes with a final \textit{Dense} output layer for classification, configured identically to the one used in the linear classifier, with \textit{CategoricalCrossentropy} as the loss function.

\subsubsection{Classification using a single pixel}\label{sec:methods_results_pixel}

The output signals of the MRR network, corresponding to the flattened output images, are downsampled from 1225 time samples (with relative experimental indices) to 409 output pixels, in order to speed up computations and reduce the effect of experimental noise and low vertical resolution of the oscilloscope (averaging smoothens the signal). This is done using the Python function \textit{skimage.measure.block\_reduce} from the \textit{scikit-image}, based on the average of time samples and with a downsample factor 3. It should be stressed that this operation approximates the use of a slower (and cheaper) photodetector, which would not introduce any additional computational cost. Moreover, to speed up the computations producing the ML accuracies reported in \cref{fig:single_pix_summary} and \cref{tab:single_pixel}, we exclude the first 80 time samples from the investigation (with the exception of the digital baseline results), as they convey little information about the image class and consistently provide low classification accuracy.
For this part of our work, we employ a multinomial logistic regression as a readout linear classifier, using the Python function \textit{sklearn.linear\_model.LogisticRegression} from the \textit{scikit-learn} library, with solver option \textit{lbfgs}.

\section*{Data and code availability statement}
The raw measured data supporting the results of this study are available from the corresponding author upon reasonable request. Other data and the code supporting the results of this study is openly available at the following URL/DOI: \url{https://doi.org/10.5281/zenodo.17105607}.

\section*{Acknowledgments}
We acknowledge fruitful discussion with Dr. Stefano Biasi.
A.F. acknowledges funding by the European Union under GA n°101070238-NEUROPULS. A.L. acknowledges funding by the European Union under GA n°101064322-ARIADNE. Views and opinions expressed are however those of the author(s) only and do not necessarily reflect those of the European Union or The European Research Executive Agency. Neither the European Union nor the granting authority can be held responsible for them.

\bibliographystyle{unsrt}
\bibliography{references}

\section*{Contributions}
A.F. and A.L. contributed equally to this work.

A.F. and A.L. conceived the experiments, prepared the setup and performed the experimental measurements. A.L. designed the integrated photonic circuit. A.F. performed the data analysis for section \ref{sec:multiport}, A.L. for section \ref{sec:single_pix}. A.F. and A.L. wrote the manuscript. P.B. and L.P supervised the work. All authors contributed to the revision of the manuscript.

\end{document}